\documentclass[twocolumn,showpacs,preprintnumbers,amsmath,amssymb]{revtex4}

\usepackage{graphicx}% Include figure files
\usepackage{dcolumn}% Align table columns on decimal point
\usepackage{bm}% bold math

\newcommand{\bra}[1]{\langle #1|}
\newcommand{\ket}[1]{|#1\rangle}

\begin{document}

%\preprint{APS/123-QED}

\title{Entanglement as a quantum order parameter}

\author{Fernando G. S. L. Brand\~ao}
\email{fgslb@ufmg.br}
\affiliation{${}^{1}$QOLS, Blackett Laboratory, Imperial College London, London SW7 2BW, UK}
\affiliation{${}^{2}$Institute for Mathematical Sciences, Imperial College London, London SW7 2BW, UK}
\affiliation{${}^{3}$Grupo Informa\c c\~ao Qu\^antica, Departamento de F\'{\i}sica - CP 702 - Universidade Federal de Minas Gerais, 30123-970, Belo Horizonte, Brasil}
\date{\today}

\begin{abstract}

We show that the quantum order parameters (QOP) associated with the transitions between a normal conductor and a superconductor in the BCS and $\eta$-pairing models and between a Mott-insulator and a superfluid in the Bose-Hubbard model are directly related to the amount of entanglement existent in the ground state of each system. This connection gives a physical meaningful interpretation to these QOP, which shows the intrinsically quantum nature of the phase transitions considered.  

\end{abstract}

\maketitle

\section*{}

It has recently become clear that entanglement plays an important role in the understanding of critical and thermodynamical properties of quantum systems \cite{ost, osb, vid, fan2, anf, ved, plen}. The theory of entanglement, developed in the context of quantum information processing, has been applied in the study of properties of several important condensed matter systems, such as spin chains \cite{ost, osb, vid, fan2, anf, ved, yu} and bosonic systems described by quadratic Hamiltonians \cite{plen}. Particularly, it was shown that near a quantum phase transition entanglement can be classified in the framework of scaling theory \cite{ost, vid} and that, using the so-called localizable entanglement, it is possible to define the concept of entanglement length which, in some cases, can detect phase transitions undetected by the traditional correlation length \cite{ver}. In this letter we add another element to this picture, studying how entanglement is related to the order parameter (a quantity that assumes non-zero values in one phase, while is strictly zero in the other) in some important quantum models.       

Classical order parameters usually have a clear physical interpretation, e.g., the density in the transition between a liquid and a gas and the magnetization in the transition of a paramagnet into a ferromagnet. On the other hand, quantum order parameters (QOP) are not normally associated with any physical meaningful quantity. We show that the order parameters associated with the BCS model for normal superconductivity, with the $\eta$ pairing model for hight temperature superconductivity and with the Bose-Hubbard model for superfluidity are directly related to the amount of entanglement presented in those systems. This intriguing connection indicates a physical interpretation for (some) QOP: they quantify how quantum correlated are the parts of the system.

\textit{The BCS Model:} The BCS model is the most successful microscopic theory to describe superconductivity at temperatures near to zero. Some properties of entanglement in the BCS model were studied in Refs. \cite{delg, dun}. In Ref. \cite{dun} in special, it was shown that entanglement is directly related to the order parameter in the reduced BCS model. We show that the this equivalence is in fact rather general for fermionic systems whose ground state is described by the BCS state.

Following Ref. \cite{ball}, consider a fermionic system described by creation operators $a_{\alpha}^{\cal y}$, where the label $\alpha$ represents the quantum numbers $\alpha = (\vec{k}, \sigma)$ of an electron and $-\alpha$ stands for the time-reversed state. The most general Hamiltonian for this type of system can be written as 
\begin{equation}
H = \sum_{\alpha,\beta}T_{\alpha \beta}a_{\alpha}^{\cal y}a_{\beta} + \frac{1}{4}\sum_{\alpha,\beta,\gamma,\delta}\bra{\alpha \beta}V\ket{\gamma \delta}a_{\alpha}^{\cal y}a_{\beta}^{\cal y}a_{\gamma}a_{\delta},
\end{equation}
where $T_{\alpha \beta}$ is the matrix element associated with the kinetical energy and any external potential, $T = P^{2}/2M + W$, and $\bra{\alpha \beta}V\ket{\gamma \delta}$ is the matrix element from the antisymmetric interaction \cite{ball}. The crucial point in the BCS theory is the construction of an Ansatz for the fundamental state of the system which takes into account the formation of Cooper pairs:    
\begin{equation} 
\ket{BCS} = \prod_{\alpha > 0}(u_{\alpha} + v_{\alpha}a_{\alpha}^{\cal y}a_{-\alpha}^{\cal y})\ket{0},
\end{equation}
where the real parameters $u_{\alpha}$ e $v_{\alpha}$ satisfy
\begin{equation}
u_{\alpha}^{2} + v_{\alpha}^{2} = 1.
\end{equation}
Is clearly seen that the BCS state consists of correlated electrons pairs, since the electrons described by $\alpha$ and $-\alpha$ are always associated. However, if either $u_{\alpha} = 1$ or $v_{\alpha} = 1$ for every $\alpha$, the ground state becomes separable, i.e., the BCS state reduces to the Haartree-Fock (HF) approximation. The criterion for superconductivity is exactly the existence of a BCS state with lower energy than the HF state. Therefore, in a qualitative way, entanglement \footnote{In the Fock Hilbert space associated with the operators $a_{\alpha}$.} is necessary for  superconductivity.  

It can be shown that the energy of the system is given by \cite{ball} 
\begin{equation} 
E_{0} = \sum_{\alpha}\frac{1}{2}(T_{\alpha \alpha} + \epsilon_{\alpha})v_{\alpha}^{2} - \frac{1}{2}\sum_{\alpha > 0}\frac{\Delta_{\alpha}^{2}}{[(\epsilon_{\alpha} - \mu)^{2} + \Delta_{\alpha}^{2}]^{1/2}},
\end{equation}
where $\mu$ is the chemical potential, 
\begin{equation}
\epsilon_{\alpha} = T_{\alpha \alpha} + \sum_{\beta}\bra{\alpha \beta}V\ket{\alpha \beta}v_{\beta}^{2},
\end{equation}
is the energy of each Cooper pair and
\begin{equation}
\frac{1}{2}\frac{\Delta_{\alpha}}{[(\epsilon_{\alpha} - \mu)^{2} + \Delta_{\alpha}^{2}]^{1/2}} = u_{\alpha}v_{\alpha}.
\end{equation}
The first term in the R.H.S. of equation (4) is just the HF energy. We thus see that the condition for the existence of superconductivity is that at least one of the $\Delta_{\alpha}$ is non-zero. The $\Delta_{\alpha}$ are considered the order parameters and have quantitative influence on the superconductivity properties of the system \cite{ball}. 

Let us now calculate the entanglement in the BCS state. We will focus on the entanglement in momentum space, i.e., entanglement between all electrons having quantum numbers $\alpha$ and their time-reversed electrons described by $-\alpha$. We use the logarithmic negativity, a proper measure of entanglement, to this aim \cite{neg}. The negativity of a quantum state $\rho$, ${\cal N}(\rho)$, is the sum of the absolute value of the negative eigenvalues of the partial transpose $\rho^{\Gamma}$. The logarithmic negativity is then given by l${\cal N}(\rho) = \log \left(1 + 2{\cal N}(\rho) \right)$ \cite{neg}.

From equation (6) is easily seen that the negativity between the electrons described by $\alpha$ and $-\alpha$ is given by ${\cal N}_{\alpha,-\alpha} = u_{\alpha}v_{\alpha}$. Using the additivity property of the logarithmic negativity \cite{neg} we then find that the entanglement between all the electrons described by $\alpha$ and the time-reversed electrons described by $-\alpha$ can be written as  
\begin{eqnarray}
E_{\cal N}(BCS) = \sum_{\alpha}\log\left(1 + \frac{\Delta_{\alpha}}{[(\epsilon_{\alpha} - \mu)^{2} + \Delta_{\alpha}^{2}]^{1/2}}\right).
\end{eqnarray}
One finds that entanglement is a monotonic increasing function of the order parameters, having, therefore, quantitative influence in the superconductivity properties of the system. Actually, the logarithmic negativity by itself could be used as an order parameter. 

%Na referência \cite{334}, o emaranhamento no modelo BCS reduzido, aplicável no estudo da supercondutividade em nano-sistemas metálicos, foi estudado. Neste caso específico, a soma das concorrências entre os elétrons descritos por $\alpha$ e $-\alpha$ é igual ao parâmetro de ordem do sistema.

\textit{The $\eta$ pairing model:} Despite the great success of the BCS theory in describing superconductivity at very low temperatures, it fails in explaining the superconductivity of some materials at high temperatures. Actually, the mechanism which allows the existence of superconductivity at temperatures as high as 160K is not yet completely understood. We now analyze a particular mechanism for $T_{C}$ superconductivity, the $\eta$ pairing of electrons \cite{yang}. The biggest difference between this and the BCS model is that in the former Coopers pairs are formed from electrons at the same site, whereas in the later electrons forming a Cooper pair have an average finite separation distance. Some of the entanglement properties of the $\eta$ pairing model were studied in Refs. \cite{zan, ved2, ved3, fan}. It was particularly shown that again in this model the existence of entanglement is a necessary condition for superconductivity \cite{ved2, ved3}. We will expand this result, establishing relations between the amount of entanglement and the order parameter of the system in both the finite and thermodynamical regimes.  

Following Ref. \cite{ved2}, consider a set of sites, where each one can be occupied by fermions having spin down and up. Let $c_{i,s}^{\cal y}$ be the fermionic creation operator, where the indices $i$ and $s$ identify the $i$-th site and the spin orientation, respectively. The operators $c$ satisfy the following commutation relations  
\begin{equation} 
{\cal f}c_{i,s}, c_{j,p}^{\cal y}{\cal g} = \delta_{ij}\delta_{sp}.
\end{equation}
The operator that creates a coherent superpositions of Cooper pairs in each of the sites, $\eta^{\cal y}$, is given by
\begin{equation}
\eta^{\cal y} = \sum_{i=1}^{n}c_{i,\uparrow}^{\cal y}c_{i,\downarrow}^{\cal y},
\end{equation}
where $n$ is the number of sites of the system. The operators $\eta^{\cal y}$ can be applied several times, where in each one a new superposition is created. Nonetheless, due to the Pauli exclusion principle, the number of applications cannot exceeds the number of sites. The state of a system in which $k$ coherent pairs were created is given by 
\begin{equation}
\ket{k, n - k} = \binom{n}{k}^{-1}(\eta^{\cal y})^{k}\ket{0}.
\end{equation}
Note that in this representation each site is described effectively by one qubit, whose value 0 stands for an empty site and 1 for a occupied one \cite{ved2}.

The most important characteristic of the $\eta$ states is the existence of off diagonal long range order (ODLRE), which implies the main superconductor properties, such as the Meissner effect and the flux quantization \cite{ved3}. The ODLRE is defined by a non-zero value of the off diagonal elements of the reduced density matrices of two sites, when the distance between these two becomes arbitrarily long: 
\begin{equation}
\lim_{|i - j|\rightarrow \infty}\left< c_{j,\uparrow}^{\cal y}c_{j,\downarrow}^{\cal y}c_{j,\downarrow}c_{j,\uparrow} \right> \rightarrow \alpha,
\end{equation}
where $\alpha$ is a constant independent of $n$ in the thermodynamical limit. The number $\alpha$ is exactly the order parameter of the high-$T_{C}$ superconductivity \cite{yang}. 

The reduced density matrix $\sigma_{ij}$ for the sites $i$ and $j$ of the state $\ket{n, n - k}$ is 
\begin{equation}
\sigma_{ij} = a\ket{00}\bra{00} + b\ket{11}\bra{11} + 2c\ket{\psi^{+}}\bra{\psi^{+}},
\end{equation} 
where $\ket{\psi^{+}} = (\ket{01} + \ket{10})/\sqrt{2}$,
\begin{equation}
a = \frac{k(k - 1)}{n(n - 1)}, \hspace{0.4 cm} b = \frac{(n - k)(n - k - 1)}{n(n - 1)},
\end{equation}
and
\begin{equation}
c = \frac{2k(n - k)}{n(n - 1)}.
\end{equation}
Since the state is symmetric, this density matrix is the same no matter how far the two sites are from each other. The order parameter is thus easily found to be $\alpha = c$. In the next paragraphs we associate it with the entanglement of the system. In this model we study the entanglement between the occupation number of each site. 

Consider first the case of finite $n$. That is the case, for instance, of superconductivity in nano-structures \cite{delft}. In Ref. \cite{ved2} it was shown that, in this case, the two sites density matrices are entangled for every $n$ and $k$, and that the order parameter $c$ is just the maximum fidelity of teleportation under local operations and classical communication \footnote{A direct expression between the order parameter and the negativity can also be straighforwardly derived.}\cite{hor}. Notably, we see that in the finite case the number $c$ quantifies both the superconductivity features of the system \textit{and} its usefulness as a quantum channel! 

For general superconducting materials we must consider the thermodynamical regime, where $n, k \rightarrow \infty$. In this scenario the analysis becomes more complex, as the states $\sigma_{ij}$ become separable no matter what are the ratio between $n$ and $k$ \cite{ved2, fan}. From equation (11) we find that the order parameter is non-zero iff
\begin{equation}
\lim_{n, k \rightarrow \infty}\frac{k}{n} = r,
\end{equation}
for some real number $r$. As noted by Vedral \cite{ved2, ved3}, although there is no bipartite entanglement between two sites in this regime, the system still has multipartite entanglement \footnote{And, in fact, bipartite entanglement between any split of the whole system in two parts.}. In order to quantify it, we will use the logarithm geometric measure of entanglement $LR_{G}$, which is given by the logarithm of the overlap between the state and its nearest separable state in the norm 2 sense \cite{wei}. This measure was calculated for the states $\ket{n, n - k}$ and is given by \cite{wei}
\begin{equation*}
LR_{G}(n, k) =  \log \left(\frac{k!(n - k)!}{n!}\left(\frac{n}{k} \right)^{k}\left(\frac{n}{n - k} \right)^{n - k}\right).
\end{equation*}
Using $k = rn$ and the Stirling approximation $ln(n!) \approx n\ln n - n$, we have
\begin{eqnarray}
LE_{G}(n, k) = \log \left[(1 + r)\left(\left(\frac{1}{r}\right)^{r}\left(\frac{1}{1 - r}\right)^{1 - r}\right)^{n}\right] \nonumber \\
+ O(\frac{1}{n}, \frac{1}{k}). \nonumber
\end{eqnarray}
One might note that when $k \rightarrow \infty$, the amount of entanglement diverges. However, if we use instead the density of entanglement, i.e., the average entanglement per site, $d_{E} = LE_{G}/n$, we find
\begin{eqnarray} 
d_{E} = \lim_{n \rightarrow \infty} \frac{LE_{G}(n, k)}{n}
= -r\log r - (1 - r)\log(1 - r) \nonumber
\end{eqnarray}
From equation (15) one finds in addition that $\alpha = 2r(1 - r)$. Therefore, it holds the following direct relation between the amount of multipartite entanglement and the order parameter
\begin{eqnarray} 
d_{E} = -\left(\frac{1 - \sqrt{1 - \alpha}}{2}\right)\log \left(\frac{1 - \sqrt{1 - \alpha}}{2}\right) \nonumber \\
- \left(\frac{1 + \sqrt{1 - \alpha}}{2}\right)\log \left(\frac{1 + \sqrt{1 - \alpha}}{2}\right).
\end{eqnarray}
Interestingly, the relation between $d_{E}$ and $\alpha$ is same relation of the entanglement of formation and the concurrence \cite{woo}, which implies that also in this model the order parameter quantifies entanglement.      

\textit{The Bose-Hubbard Model:} The simplest non-trivial model for interacting bosons in a periodic potential is described by the Bose-Hubbard (BH) Hamiltonian \cite{sach}. It contains most of the physics of strongly interacting bosons, i.e., the competition between kinetic and interaction energy. The BH Hamiltonian is given by \cite{sach}
\begin{equation}
\hat{H}_{BH} = -J\sum_{<n,m>}a_{n}^{\cal y}a_{m} + \frac{U}{2}\sum_{n}a_{n}^{\cal y}a_{n}^{\cal y}a_{n}a_{n} + \sum_{n}(V_{n} - \mu)a_{n}^{\cal y}a_{n},
\end{equation}
where $a_{m}$ is the annihilation operator for an atom at site $m$. The first term in equation (17), proportional to $J$, is the tunneling matrix element between nearest-neighbors. The parameter $U$ is proportional the repulsion intensity of two atoms at the same site.  

A zero temperature, the physics of the Bose-Hubbard model can be divided in to extreme regimes. The first is the one where $J$ is much lower than $U$ and the system is described by a Mott insulator. The other is the regime dominated by the kinetic energy, where $J$ is much larger than $U$ and the system presents superfluidity properties \cite{sach}.

The ground state of the Bose-Hubbard Hamiltonian cannot be found analytically for every $U$ and $J$. Therefore, we will restrict our analysis to the limiting cases $U/J \rightarrow \infty$ and $U/J \rightarrow 0$. Consider a system with $N$ atoms. When $U/J \rightarrow 0$, its ground state is a deeply superfluid state given by \cite{sach} 
\begin{equation}   
\ket{\psi_{SF}} = \frac{1}{\sqrt{N!}}\left(\frac{1}{\sqrt{M}}\sum_{m=1}^{M}a_{m}^{\cal y}\right)^{N}\ket{0},  
\end{equation}   
where $M$ is the number of wells of the system.  In the other limit, where the system is described by a Mott insulator, the ground state is just a separable state given by
\begin{equation}   
\ket{\psi_{IM}} = \prod_{i=1}^{M}\ket{g_{i}},
\end{equation}   
where $\ket{g_{i}}$ are variational local states. The order parameter usually considered in this model is the expectation value of the annihilation operator $a_{m}$ \footnote{As the Bose-Hubbard Hamiltonian is translational symmetric, $\left<a_{m}\right>$ is the same for all $m$.}. In the thermodynamical limit $N, M \rightarrow \infty$, with 
\begin{equation}
\lim_{N, M \rightarrow \infty} \sqrt{\frac{N}{M}} = r,
\end{equation}
it can be shown that in the MI phase $\left<a_{m}\right> = 0$, whereas in the superfluidity phase $\left<a_{m}\right> = r$ \cite{sach}. 

Consider now the bipartition where one party is formed by the Fock space associated with the operator $a_{m}$ and the other by the Fock space associated with all other operators $a_{n}$, $n \neq m$. The state $\ket{\psi_{SF}}$ can then be written as
\begin{equation}
\ket{\psi_{SF}} = \frac{1}{N!}\left(\sqrt{\frac{1}{M}}a_{m}^{\cal y} + \sqrt{1 - \frac{1}{M}}b^{\cal y}\right)^{N}\ket{0},
\end{equation}
with $b = \sqrt{\frac{1}{M - 1}}\sum_{n \neq m}a_{n}$. Equation (21) can expressed in terms of its Schmidt decomposition
\begin{equation}
\ket{\psi_{SF}} = \sum_{k = 0}^{N}\sqrt{\binom{N}{k}}p^{\frac{k}{2}}(1 - p)^{\frac{N - k}{2}}\ket{k}_{a_{m}}\ket{N - k}_{b},
\end{equation}
where $p = 1/M$. Let us calculate the entanglement in this partition, which quantifies the quantum correlations existent between the number of atoms in site $m$ and the number of atoms on all the other sites. Using the negativity \cite{neg} as a measure of entanglement,
\begin{equation}
{\cal N}(\psi_{SF}) = \frac{\left( \sum_{k = 0}^{N}\sqrt{\binom{N}{k}p^{k}(1 - p)^{N - k}}\right)^{2} - 1}{2}.
\end{equation}
From the central limit theorem we have that in the limit $N \rightarrow \infty$,
\begin{equation}
\binom{N}{k}p^{k}(1 - p)^{N - k} \rightarrow \frac{1}{\sqrt{2\pi Np(1 - p)}}e^{\frac{(k - Np)^{2}}{2Np(1 - p)}}.
\end{equation}
Thus we can replace the sum in equation (24) by the integral
\begin{equation}
\frac{1}{(2\pi Np(1 - p))^{1/4}}\int_{k=0}^{\infty}dk e^{\frac{(k - Np)^{2}}{4Np(1 - p)}} = (8Np(1 - p))^{1/4}.
\end{equation}
Therefore, in this limit, we have again a direct relation between entanglement and the value of the order parameter:
\begin{eqnarray}
\lim_{N, M \rightarrow \infty} {\cal N}(\psi_{SF}) = \lim_{N, M \rightarrow \infty} \left(2\frac{N}{M}\left(1 - \frac{1}{M}\right) \right)^{1/2} - \frac{1}{2} \nonumber \\ 
= \sqrt{2}r - \frac{1}{2}. \nonumber
\end{eqnarray}

%\textit{The Higgs-Mechanism:} One of the most elegant manner to explain massive particles in the standard model is the through the Higgs mechanism \cite{wein}, where the condensation of the Higgs field, in a complete analogy with the meissner effect, expels the other fields from a determined region. In that way, these other fields become short ranged and their mediators massive. As it was recently argued by Vedral \cite{ved3}, if the Higgs mechanism indeed happens, then its mediators particles, the Higgs bosons, must be entangled. The condensation of the Higgs bosons can be considered a quantum phase transition, where the order parameter is just the mass of the mediators particle. If the quantitative relation between entanglement and quantum order parameters is also valid in these case and the theory of entanglement is extended so that it is able to determine the entanglement content of quantum fields, then we would have a mean to calculate the masses of the particles in the standard model! 

In conclusion, we have shown that there exist a direct relation between entanglement and the order parameter in three important physical systems. This connection has important consequences both in the theory of phase transitions and in quantum information science. On one hand, quantum order parameters gain an interesting physical interpretation, which clarifies the intrinsic quantum character of QFT. On the other hand, the rich literature about the properties of QOPs, including evolution equations and statistical properties, can be used in the study of entanglement and the viability of quantum information processing in those systems.

\textit{Note added:} After this work was completed, I learned that a connection between single-site entanglement and the energy-gap in the BCS model was obtained by Shi in Ref. \cite{yu}. In this same paper an interesting relation between entanglement and the proper fractional part of the filling factor is also derived, which extends the discussion of the present paper also to the quantum Hall state.

The author would like to thank Jens Eisert and Martin Plenio for helpful comments on the manuscript and Conselho Nacional de Desenvolvimento Cient\'ifico e Tecnol\'ogico (CNPq) for financial support.

\end{document}